# Intelligent Traffic Light Control Using Distributed Multi-agent Q Learning

Ying Liu[1,2], Lei Liu[1], Wei-Peng Chen[1]
[1] Fujitsu Laboratories of America, Inc., Sunnyvale, CA, USA
[2] Electrical and Computer Engineering, Rutgers University, New Brunswick, NJ, USA

***Abstract*—The combination of Artificial Intelligence (AI) and Internet-of-Things (IoT), which is denoted as AI powered Internet-of-Things (AIoT), is capable of processing huge amount of data generated from large number of devices and handling complex problems in social infrastructures. As AI and IoT technologies are becoming mature, in this paper, we propose to apply AIoT technologies for traffic light control, which is an essential component for intelligent transportation system, to improve the efficiency of smart city's road system. Specifically, various sensors such as surveillance cameras provide real-time information for intelligent traffic light control system to observe the states of both motorized traffic and non-motorized traffic. In this paper, we propose an intelligent traffic light control solution by using distributed multi-agent Q learning, considering the traffic information at the neighboring intersections as well as local motorized and non-motorized traffic, to improve the overall performance of the entire control system. By using the proposed multi-agent Q learning algorithm, our solution is targeting to optimize both the motorized and non-motorized traffic. In addition, we considered many constraints / rules for traffic light control in the real world, and integrate these constraints in the learning algorithm, which can facilitate the proposed solution to be deployed in real operational scenarios. We conducted numerical simulations for a real-world map with real-world traffic data. The simulation results show that our proposed solution outperforms existing solutions in terms of vehicle and pedestrian queue lengths, waiting time at intersections, and many other key performance metrics.**

## I. Introduction

In the upcoming Internet-of-Things (IoT) era, there will be many complex systems, networks or social infrastructures, connecting huge number of devices which will generate huge amount of data. To manage these devices and data, as well as provide intelligent control in such complex scenarios, artificial intelligence (AI), for its capability and potential for handling complicated tasks, has been widely investigated in recent years. With the maturity of AI and IoT, it is essential to evaluate the feasibility and efficiency of AI based intelligent transportation system (ITS) because the ITS is closely related to people's daily life and is one important aspect of building a smart city. According to a recent competition report [1] made by U.S. Department of Transportation, a lot of cities have a common challenge to manage the traffic flow and reduce the traffic congestion. Considering the fact that the traffic light control system is one of the most straightforward approaches to address the above challenge, in recent years, many solutions and algorithms for traffic light control have been proposed in order to improve the traffic congestion.

However, most previous solutions and algorithms only focused on motorized traffic (e.g. vehicles). Non-motorized traffic (such as pedestrian and bicyclist) is rarely considered, and in most cases, non-motorized users have to manually activate the timing system by pushing a button, which may affect the overall efficiency of the entire traffic light control system. Therefore, to improve the overall efficiency of traffic light control, dynamic coordination of vehicular traffic with non-motorized traffic should be taken into account. However, modeling the correlation between vehicular and pedestrian's mobility is a challenging task since it is influenced by many factors which are uncertain and dynamic with time. Any preset control mechanism which is based on a particular rule may not be able to address such a dynamic problem.

In this paper, by leveraging the notion of AI and IoT, we propose an intelligent traffic light control solution by using distributed multi-agent Q learning. The solution considers not only motorized traffic, but also non-motorized traffic, by dynamically monitoring and collecting vehicle and pedestrian queue lengths at each intersection. The proposed Q learning algorithm is implemented at each intersection, where the Q learning agent interacts with environment to learn the optimal control actions to minimize the length of waiting queues for both vehicle and pedestrian traffic. The observations at individual intersection are exchanged with its neighboring intersections through the network in a distributed manner to achieve the global optimal schedule for the entire system. Moreover, we considered many constraints / rules for traffic light control in the real world, and integrate them in the learning algorithm, which can facilicate the proposed solution to be deployed in real operational scenarios. To validate the efficiency of the proposed solution, we conducted numerical simulations by using a real-world map (from the OpenStreetMap [2], [3]) and real-world traffic data (from California Department of Transportation [4]). The simulation results show that our proposed solution outperforms existing solutions in terms of vehicle and pedestrian queue lengths, waiting time at intersections, and many other performance metrics.

The remainder of this paper is structured as follows: In Section II we briefly review some related works. In Section III a brief introduction for single agent and multi-agent Q-learning is presented. In Section IV, we illustrate details for our proposed traffic light system and algorithms. Section V shows simulation results by using real world map and traffic data. Finally, we summarize the paper and present our future works in Section VI.

## II. Related works

Traffic lights are signalling devices which have been widely deployed in the world at road intersections, pedestrian crossings, and other locations to control traffic flows. Traffic lights can greatly affect the traffic condition. A well-designed control algorithm can increase the traffic handling capacity of roads, reduce collisions, waiting and traveling time for both vehicles and pedestrians. On the contrary, an inefficient control algorithm may cause significant traffic congestion, resulting in longer waiting time at intersections. To this end, various traffic signal control techniques have been proposed in the recent years.

The commonly used one is fixed-time control where traffic signals are changed after a fixed time period (aka. threshold). However, this threshold can be pre-configured to different values based on different time periods in a day. The other widely deployed solution is dynamic control. To support this solution, detectors such as sensors or surveillance cameras [5] are deployed in the intersections to detect whether vehicles are present or not. The traffic light control module, based on this information, can adjust signal timing and phasing within the pre-determined limits. Moreover, there are some advanced solutions proposed in recent years, which are referred to as adaptive control. The adaptive control solutions are usually associated with centralized or distributed coordination among multiple intersections, and try to change or adapt traffic signal timing based on actual traffic demand. Although the adaptive traffic control is superior to the fixed-time and dynamic control, adaptive traffic light control systems have been deployed on less than one percent of existing traffic signals according to a recent report [6]. Some of the existing systems of adaptive traffic signal include Split Cycle Offset Optimization Technique (SCOOT), Sydney Coordinated Adaptive Traffic System (SCATS) which are both centralized, and Real Time Hierarchical Optimized Distributed Effective System (RHODES) which is decentralized with complex computation. The report [7] compares the performance of these solutions, and investigates their advantages and disadvantages.

Considering the challenges for modeling the correlation between vehicular and pedestrian's mobility for its uncertain and dynamic nature, machine learning which arises recently is applied to traffic light control by researchers and shows proven performance [8]–[11]. Among all the machine learning algorithms, reinforcement learning or Q learning, due to its advantage of making decision in a model-free online fashion, has been adopted by several works to design a traffic light system. The work [12] applies a Q-learning and neural network method to decide green light periods in each intersection based on traffic information. However, each intersection only calculates based on its local information, and only tries to optimize local performance. The other work in [8] applies multi-agent reinforcement learning, but each Q learning agent works independently and has not considered other intersections' status. Thus, it is hard to achive the global optimization.

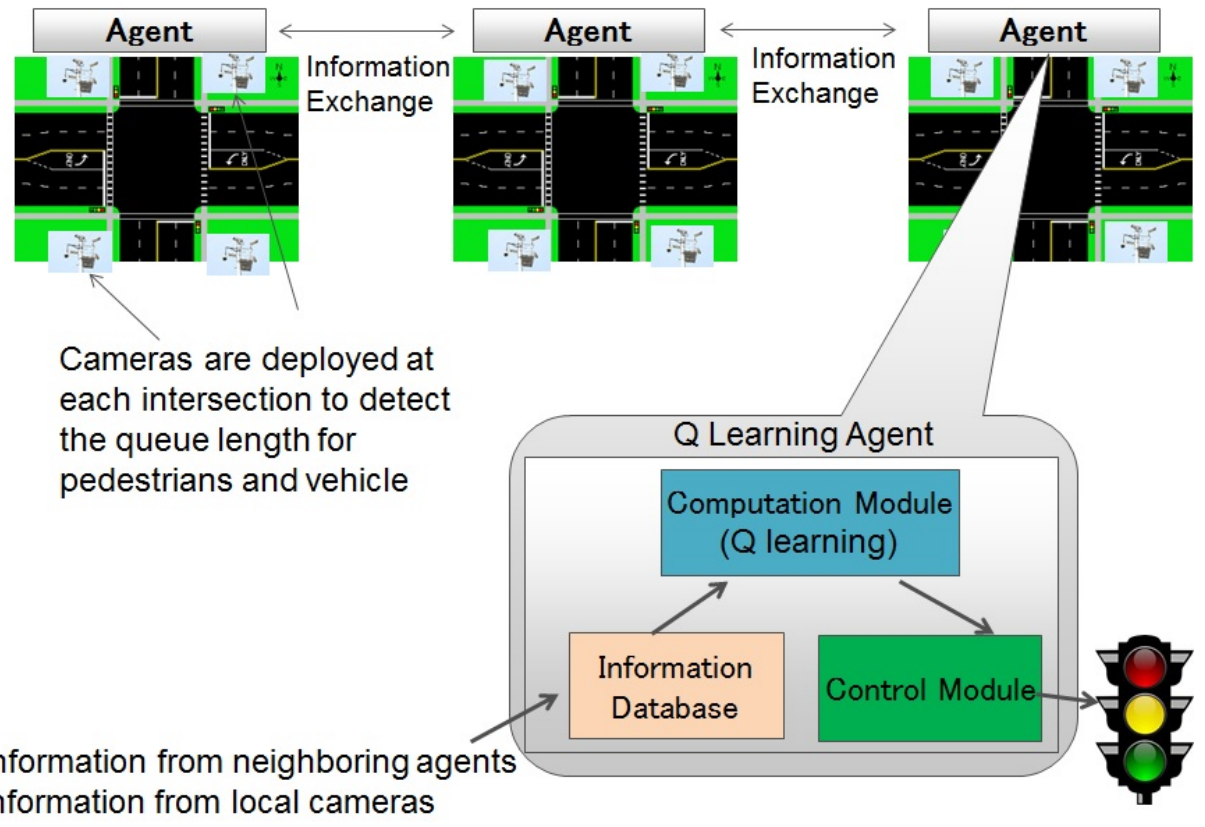

Fig. 1: System architecture for distributed multi-agent Q learning

Although the traffic light control solutions have been widely studied, most of the previously works haven't considered non-motorized traffic. Pedestrians and bicyclists have to manually activate the timing system by pushing a button, affecting the overall efficiency of the traffic light control algorithms. Only a limited number of exceptions, for example, the work [13] considers pedestrian crossing in their traffic light control by using a genetic algorithm. In this algorithm, pedestrian metric is expressed in fitness function to evaluate effectiveness of candidate chromosome. However, this work only considers one intersection for local optimization only. Compared with the previous works, the novelties and contributions of this work include: 1) the proposed solution takes both the motorized and non-motorized traffic into consideration; 2) the proposed considers not only local information but also neighboring intersections for global optimization; and 3) multiple real-world constraints and traffic rules are included in the proposed algorithm, which can be flexibly extended and easily applied to real operational scenarios.

## III. Preliminary

Reinforcement learning [14], which is an online learning method, assists an agent to take a series of optimal actions to the environment, then obtains an instantaneous reward to maximize the cumulative benefit over time. The agent's knowledge is reinforced during the learning process. The idea of reinforcement learning has been widely used in robot control, advertising, stock investment and game theory. Q-learning, as one type of reinforcement learning, gains its popularity due to no requirement of knowledge for transition probabilities in Markov decision process (MDP). Due to this reason, it is a model-free learning. For one agent control, Q-learning is expressed as follows:

$$Q^t(s,a) = (1-\alpha)Q^{t-1}(s,a) + \alpha(R^t + \gamma \max_a Q^{t-1}(s,a))$$

where $\alpha$ is learning rate, $\gamma$ is discounted. $R^t$ is instantaneous reward at time $t$. Regarding the exploitation and exploration in Q learning, there are many existing strategies, and the most commonly used three strategies are: 1) $\varepsilon$ greedy selection, 2) Boltzmann exploration which chooses an action proportional to the probability, and 3) action selection based on tightening upper confidence bound (UCB) in [15] which adjusts the rank of Q values for action and state pairs.

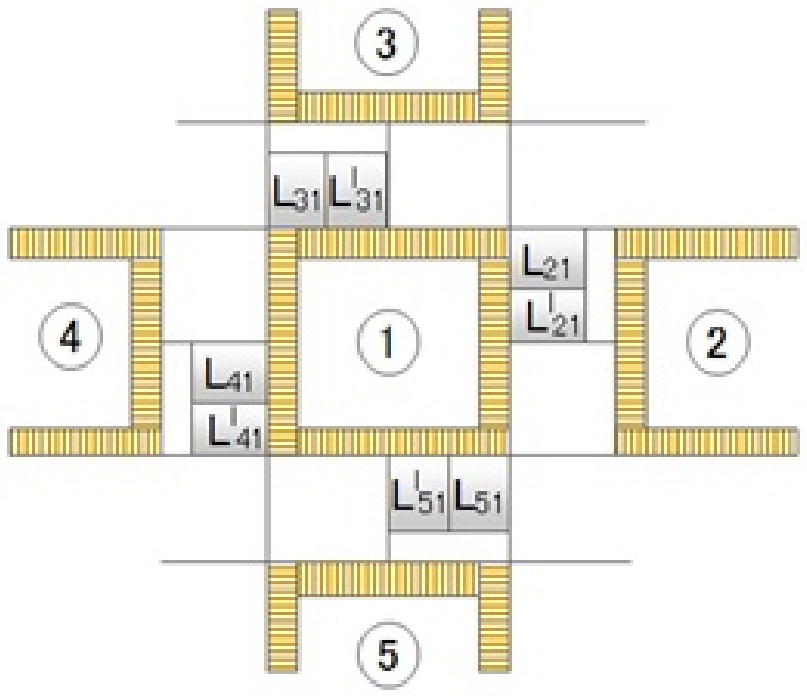

(a)

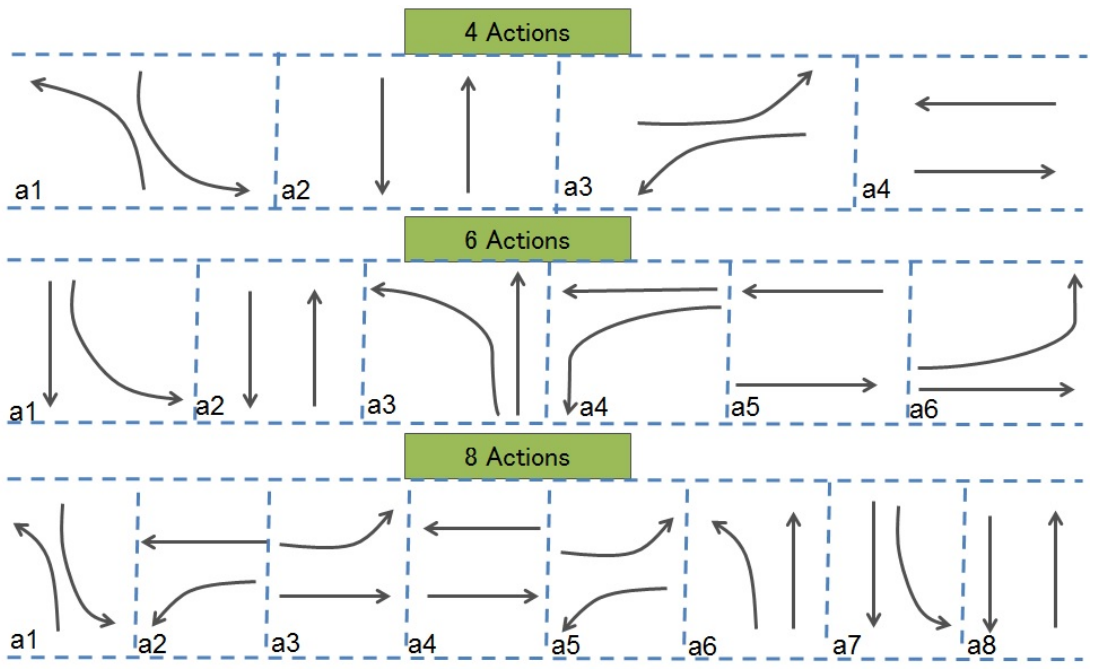

(b)

Fig. 2: Action sets (green lights) for "+" shape intersection

In a network, system or social infrastucture, an Q learning agent can be deployed on top of each entity or node, and executes Q-learning computation in a distributed manner. The agent can either work independently to optimize the local cost or collaborate with other agents (i.e. multi-agent Q learning). In the multi-agent Q learning, neighboring information can be exchanged through network connections. By exchanging messages through one hop connections, information that is multi-hop away can finally propagate across the network to achieve approximate global optimization. The collaboration among multiple agents is important since the advantage obtained by working cooperatively is usually more significant than that obtained by working independently. For transportation system, such kind of multi-agent cooperation is essentially important becuase it is not helpful by just improving traffic locally and moving congestions to different intersections.

## IV. System and algorithm design

### A. System architecture

The system architecture is shown in Fig. 1. The proposed distributed multi-agent Q-learning, as will be detailed next, is deployed in the traffic light system. At each intersection, an Q learning agent is deployed to control local traffic lights including both vehicles and pedestrians' lights in all directions. Surveillance cameras are deployed for each direction to detect queue lengths of pedestrians and vehicles. In this work, we consider the queue length is the actual number of vehicles and pedestrians waiting at the intersections for passing through. An agent collects local traffic data that are monitored by these surveillance cameras and stores the data into a local information database. Agents also collects neighborhood data by exchanging information through available network connections. These neighborhood data are also stored in the information database. Based on the database, Q computation module calculates an optimal control action, which is, in turn, executed by a control module which is a pre-programmed hardware controlling traffic light timing, as depicted in Fig. 1.

### B. State

For the state design for the proposed Q learning algorithm, we use length of waiting queues to be states for both vehicles and pedestrians in each traffic direction at an intersection, as expressed by:

$$S^t_{i,d} = \{q^t_{1i,d}, q^t_{2i,d}, ..., q^t_{ji,d}, m^t_{1i,d,L}, m^t_{1i,d,R}, ..., m^t_{ji,d,L}, m^t_{ji,d,R}\}$$

where $i, j$ are IDs of intersections and $j\epsilon N_i$; $N_i$ is neighborhood intersections of $i$; $S^t_{i,d}$ is the state at intersection $i$, at day $d$ and time $t$; $q^t_{ji,d}$ is the queue length from intersection $j$ to intersection $i$, at day $d$ and time $t$; $m^t_{ji,d,L}$ is the queue length for pedestrians at the left side from intersection $j$ to $i$, at day $d$ and time $t$; and $m^t_{ji,d,R}$ is the queue length for pedestrians at the right side from intersection $j$ to $i$, at day $d$ and time $t$. In this work, as aforementioned, we assume the surveillance cameras are deployed in the intersections to detect vehicles and pedestrians [5]. However, it should be noted that such state data can be also obtained by using other approaches and different type of detectors such as sensors, crosswalk buttons, Fastack/EasyPass, cell phones and GPS. For a large intersection which contains several straight and left lanes, state tuple could be shortened to the following expression to reduce the impact of dimensionality:

$$S^t_{i,d} = \prod_{b\epsilon D} \{q^{tb}_{r,i,d}, q^{tb}_{s,i,d}, q^{tb}_{li,d}, m^{tb}_{i,d}\}$$

Where $\prod$ means concatenate the set in $\{\}$. $D$ is a directional set which contains all directions such as north, south, east, and west. $r$ means right turn. $s$ means going straight. $l$ means left turn. $q^{tb}_{r,i,d}$ and $q^{tb}_{s,i,d}$ can be combined if there is no right turn traffic light. Then, Q-learning maintains a state-action table to keep track of old Q values for new Q value computation.

### C. Action

We design actions for our Q-learning algorithm according to current traffic light rules. In each time slot, only one action can be executed. This action is calculated and selected from the action sets by the Q learning agent which can maximize the rewards (as will be detailed next). The action sets could be different for each intersection and they are configured based on the general practice. For example, for the most common "+" shape intersections, possible action sets are shown in Fig. 2, which include 4, 6, and 8 actions respectively. Note that the "+" shape intersection and the actions sets depicted in Fig. 2 are only examples to illustrate our design, and our solution can be applied to any shape of intersections.

### D. Reward

Utilizing different rewards of Q-learning achieves corresponding different control or optimization purposes. For example, a reward can be the negative value of vehicle and non-motorized queue lengths, emission utility or traffic flows at crossing, etc. In this paper, the objective function is to minimize total average of queue length of both motorized and non-motorized traffic which is a metric that directly reflects congestion conditions. In this sense, we design the local instantaneous reward as follows:

$$R_{i,d}^t(a_{i,d}^t, a_{j,d}^t, S_{i,d}^t, S_{j,d}^t, W_{i,d}^t) = -(\frac{w_{1,d}^t}{|N_i|}\sum_{j\epsilon N_i} q_{ji,d}^t + \frac{w_{2,d}^t}{|N_i N_j|}\sum_{j\epsilon N_i}\sum_{k\epsilon N_j} q_{kj,d}^t + \frac{w_{3,d}^t}{2\,|N_i|}\sum_{j\epsilon N_i}(m_{ji,d,L}^t + m_{ji,d,R}^t)) \quad (1)$$

where $R_{i,d}^t$ is the reward at intersection $i$, at day $d$ and time $t$; $a_{i,d}^t$ is the action at intersection $i$, at day $d$ and time $t$; $w_{1,d}^t$ is the weight to present the local vehicular queues at intersection $i$; $w_{2,d}^t$ is the weight to present the neighborhood vehicular queues at the neighbors of intersection i; $w_{3,d}^t$ is the weight to present the total pedestrian queues at intersection $i$. $W_d^t = \left\{w_{1,d}^t, w_{2,d}^t, w_{3,d}^t\right\}$. The sum of these weights equals to 1.

In (1), $\sum_{j\epsilon N_i} q_{ji,d}^t$ is the incoming vehicular queues from intersection $j$ to intersection $i$; $\frac{1}{|N_j|}\sum_{j\epsilon N_i}\sum_{k\epsilon N_j} q_{kj,d}^t$ is the total vehicular queues at all neighboring intersection $j$s, including the outgoing vehicular traffic from intersection $i$ to intersection $j$; and $\sum_{k\epsilon N_j} q_{kj,d}^t$ is the total pedestrian queues at intersection $i$. When exchanging observations, intersection $j$ encapsulates the information of $\frac{1}{N_j}\sum_{k\epsilon N_j} q_{kj,d}^t$ to a status update message and broadcast it to all neighboring intersections. $|N_i|$ is decided by the shape of intersections. For example, $|N_i|$ is 4 for a four-way, "+" shaped intersection, and $|N_i|$ is 3 for a three-way, "T" shaped intersection. In (1), the weights, $w_{1,d}^t$, $w_{2,d}^t$, $w_{3,d}^t$ correlate to the priority assigned to the additive term. That is, the higher the priority, the higher the weight.

Since the reward $R_{i,d}^t$ is expressed as the negative value of queue lengths at intersection $i$, at day $d$ and time $t$, accordingly, the objective of Q-learning is to maximize the reward. In addition, due to neighboring traffic condition is considered in (1), as a result, for example, if neighboring intersections are congested, the action that causes more jams to the neighbors is less likely to be selected due to the smaller $R_{i,d}^t$. Thus, the traffic condition across multiple intersections can be achieved.

Moreover, historical traffic data can be incorporated in the determination of actions. For example, data that are days away can be used to calculate estimated instantaneous rewards by autoregressive integrated moving average (ARIMA) model [16] because traffic data in the same time period at different days may have similar statistics and patterns.

### E. Integrate real-world constraints and traffic rules

In a traffic light control system, there are a lot of rules or real-world constraints. The proposed Q learning solution takes these rules and constraints into account, which can facilicate its deployment in real operational scenarios. In this sub-section, two important constraints are presented.

*1) Constraint for action selection:* when Q-learning chooses an action from the action set for a given time slot, the selection of the next action is constrained by traffic rules. In the current real-world traffic light systems, the sequence of the actions follow some particular rules and it is unrealistic to freely choose an action. For example, in the 8 actions in Fig. 2, the constraints on the action selection could be: $a_1$ is not directly followed by $a_5$, $a_6$, $a_7$, $a_8$; $a_5$ is not directly followed by $a_1$, $a_2$, $a_3$, $a_4$; $a_2$ and $a_3$ are not directly followed by $a_5$; and $a_5$ and $a_7$ are not directly followed by $a_1$. To satisfy these constraints in Q-learning operations, we can either greedily choose another action that satisfies these conditions; or for the given action, we can assign it with a very small reward value so that Q-learning rank it low in the priority queue.

*2) Constraint for pedestrian protection:* another constraint is the pedestrian protection which is related to the safety of pedestrians. The constraint is that a traffic signal that is directing pedestrians should not turn red while pedestrians are crossing the intersection. Our solution to address this constraint is to set a short duration for the default pedestrian green light. However, if pedestrians are crossing the intersection, the Q learning will not change the control actions, which is in turn, can protect crossing pedestrians. The presence of a pedestrian can be detected by surveillance cameras at the intersections which identify walking people. On the other hand, if there is no pedestrian detected, such protection will not be enabled. Moreover, we also set a sufficient time interval for a pedestrian safely crosses an intersection if a pedestrian button is pressed in the previous red light time cycle. In this case, system allocates at least 13 time slots to the pedestrians in the next green cycle to guarantee pedestrians can finish crossing.

The pseudo code of traffic light control algorithm is shown in Algorithm 1. We have integrated the above constraints when choosing a proper action. In the pseudo code, $N(a)$ is the occurrence of action $a$. Eq. (2) represents an action selection based on USB criterion in [15]. According to Algorithm 1, the actual action selection only happens when there is no pedestrian crossing an intersection and when Q-learning computation module is scheduled another action.

## V. Performance Evaluation

We implement our algorithm of traffic light control in the Simulation of Urban MObility (SUMO) [17] which can model

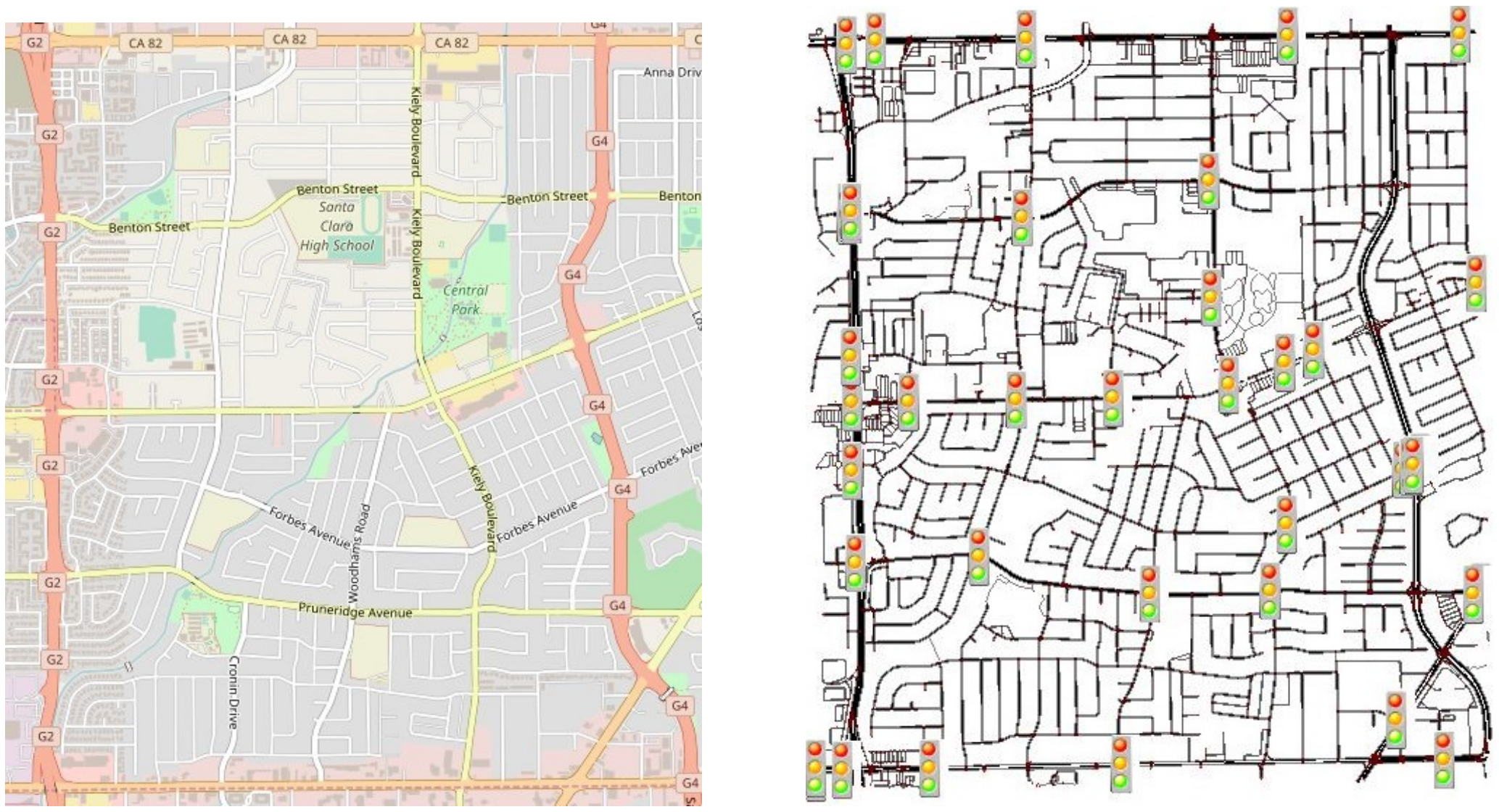

(a) (b)

Fig. 3: (a) A map of Sunnyvale, CA downloaded from OpenStreetMap; (b) the topology of the map converted by SUMO

**Data**: both motorized and non-motorized traffic data
**Result**: get optimal green time of traffic light
Initialization: $Q^0_{i,d} = 0$, action set $|A|$, $0 < \gamma < 1$, $0 < \alpha < 1$, weights $w^0_{1,d}$, $w^0_{2,d}$, $w^0_{3,d}$
randomly choose an action, $a^0_{i,d}$, from action set $A$
**for** $t$ **do**
  observe queue length in all directions, and get $q^t_{ji,d}$, $m^t_{ji,d*}$
  broadcast $T^t_{j,d} = \frac{1}{N_j}\sum_{k\epsilon N_j} q^t_{kj,d}$ and $a^t_{j,d}$ to all its neighbors
  calculate reward according to (1)
  update $Q^{t+1}_{i,d}(a^t_{i,d})$
  **if** *(execute action == Y) and (no pedestrian == Y)* **then**
    /∗ exploitation and exploration ∗/
    /∗ using either $\varepsilon$-greedy exploration strategy, Boltzmann exploration strategy or UCB ∗/
    **do**
      /∗ in the case of using UCB ∗/

$$a^{t+1}_{i,d} = arg\max_{a\epsilon A} Q^{t+1}_{i,d}(a) + \sqrt{\frac{2\log t}{N(a)}} \quad (2)$$

    **while** action constraint in Section IV-E is not met;
  **end**
**end**

**Algorithm 1:** Traffic light control

microscopic traffic conditions and has a well-designed API for controlling status of traffic light through online interaction. We export the map of an area in Sunnyvale, CA from OpenStreetMap [2], which is a rectangle area that has longitude between -121.964019 and 121.997997, and latitude between 37.322300, and 37.353056 shown in Fig. 3a. The map is converted into a SUMO compliant network topology illustrated in Fig. 3b by the netconvert tool. After convertion, the topology has a total number of 3811 edges, and 33 intersections with traffic lights. We deploy our proposed algorithm into each traffic light in Fig. 3b. In addition to the real world map, we also used real world traffic data which are obtained from the California Department of Transportation [4]. As most of the traffic statistics are related to freeways, in order to estimate the traffic for the simulation area, we firstly obtain the traffic statistics for the freeways surrounding the simulation area, and then we calculate the proportional traffic load for the simulation area for each hour in a day. These estimated values are used to model the vehicle arrival rate in the simulation. On the other hand, as there are no appropriate open data regarding pedestrians, and considering the fact that different cities may have completely different load for pedestrian traffic, in the simulation, we model 3 different arrival rates for pedestrians, which are referred to as high pedestrian rate, medium pedestrian rate, and low pedestrian rate respectively. For those 3 scenarios, the ratio between pedestrian arrival rate and vehicle arrival rate is set to 1:1, 3:5, and 1:10 respectively. The routes for both vehicle and pedestrain traffic are randomly generated by the $randomTrips$ module provided by SUMO.

For the configuration of the proposed multi-agent Q learning algorithm, we set equal weights for pedestrian, vehicle, and neighboring queues, that is, $w^t_{h,d} = 1/3$ where $h = 1, 2, 3$. Ovreall simulation time is set to be $5400$ timeslots. Initial learning rate, $\alpha$ is set to be $0.5$, and gradually decreases with time. Discounted factor, $\gamma$ is set to be 0.5. In the simulation, Q-learning does exploration using the $\varepsilon$-greedy exploration strategy, with $\varepsilon$ equals to 0.3.

In order to evaluate the overall efficiency of the proposed multi-agent Q learning solution, we compare our proposed algorithm with four real-world solutions and one state-of-the-art mechine learning based solution for traffic light control. The real-world solutions, as aforementioned in section II, include: 1) fixed time control with four actions, 2) dynamic control with four actions, 3) dynamic control with six actions, and 4) dynamic control with eight actions, which can represent most cases in reality. Here, we assume sensors are used to detect vehicles and trigger corresponding dynamic control. However, the simulation results are also applicable to other detection methods as mentioned in section II. The examples of four, six and eight action sets are illustrated in Fig. 2. Note that, these real-world solutions haven't considered pedestrians and they need to manually press a button to activate the timing system. For example, in the case of dynamic control with no pedestrian pushing the button, a traffic light may turn to red if no vehicles are waiting in the green light direction, and it may turn to green in a direction where at least one vehicle is waiting. The green traffic light in each direction is programmed with a default sequence which follows a state diagram if all four directions have waiting vehicles. Furthermore, we also compare our proposed solution with the state-of-the-art mechine learning solution for traffic light control [8]. In [8], a multi-agent Q learning algorithm, which is referred to as MARL, is proposed to improve the traffic light performance. Similarly, it also does not consider the pedestrian traffic for Q learning computation.

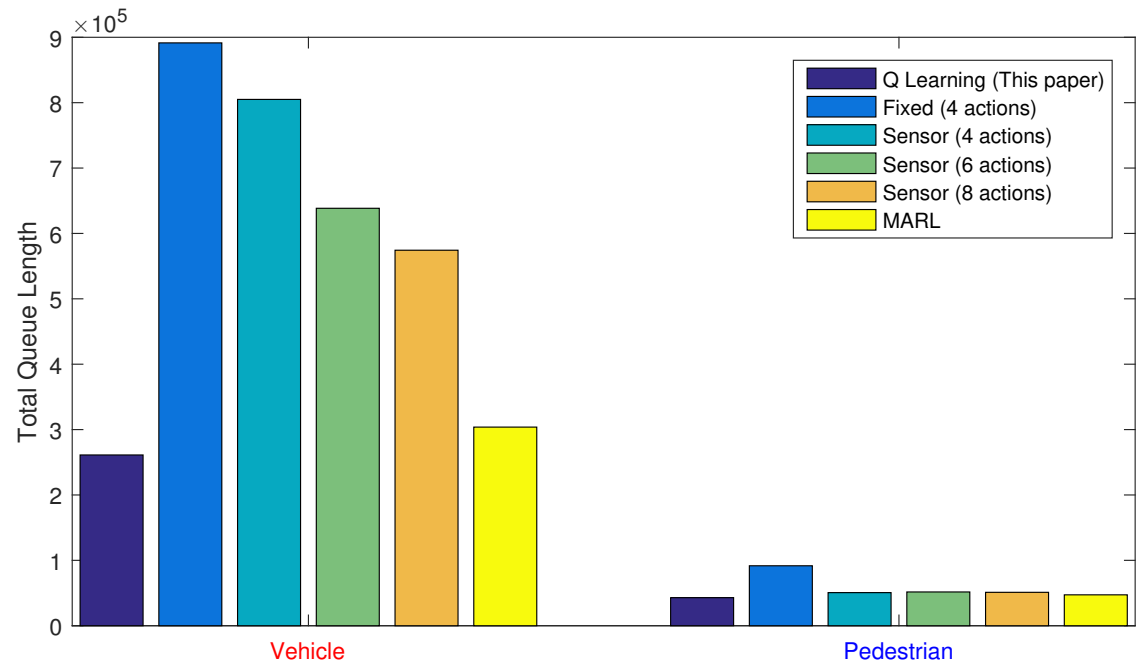

Fig. 4: Total cumulated queue lengths for vehicles and pedestrians in the case of high pedestrian rate

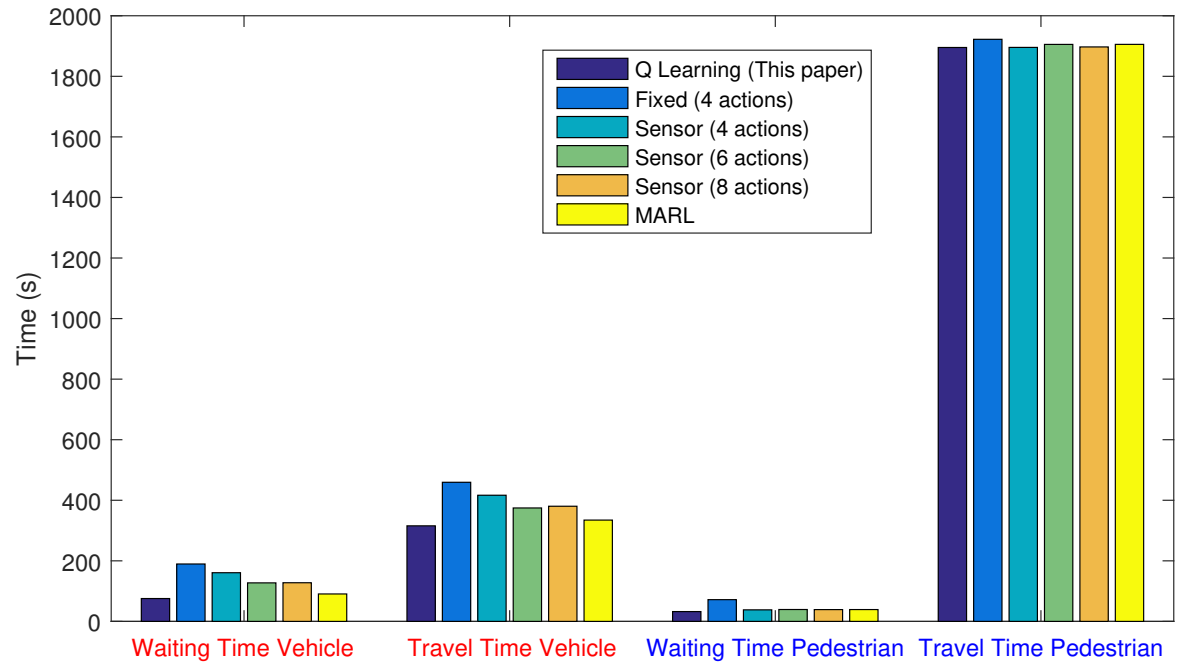

Fig. 5: Total waiting and traveling time for vehicles and pedestrians in the case of high pedestrian rate

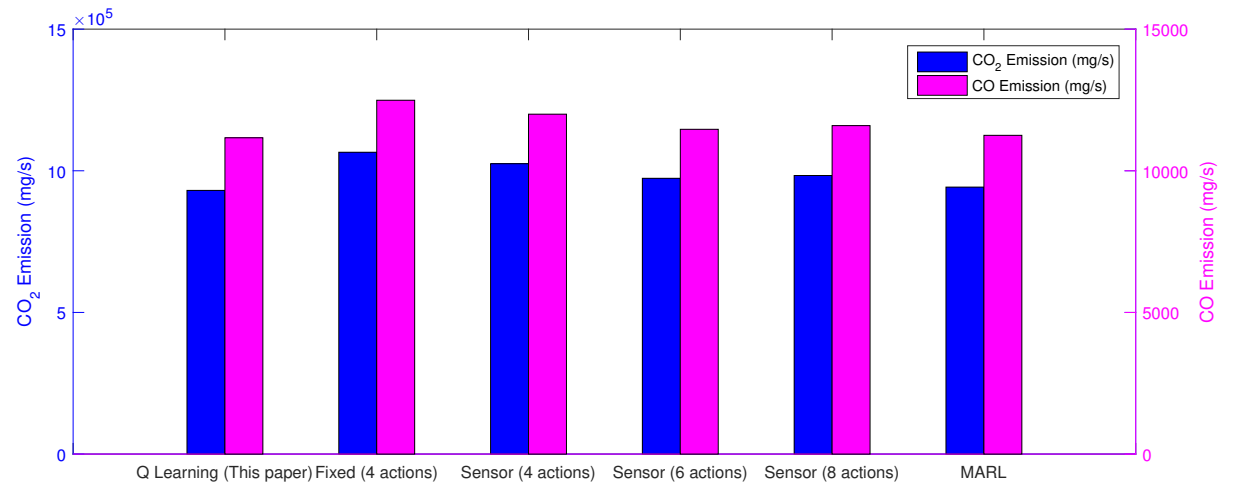

Fig. 6: $CO_2$ and $CO$ emission of vehicles in the case of high pedestrian rate

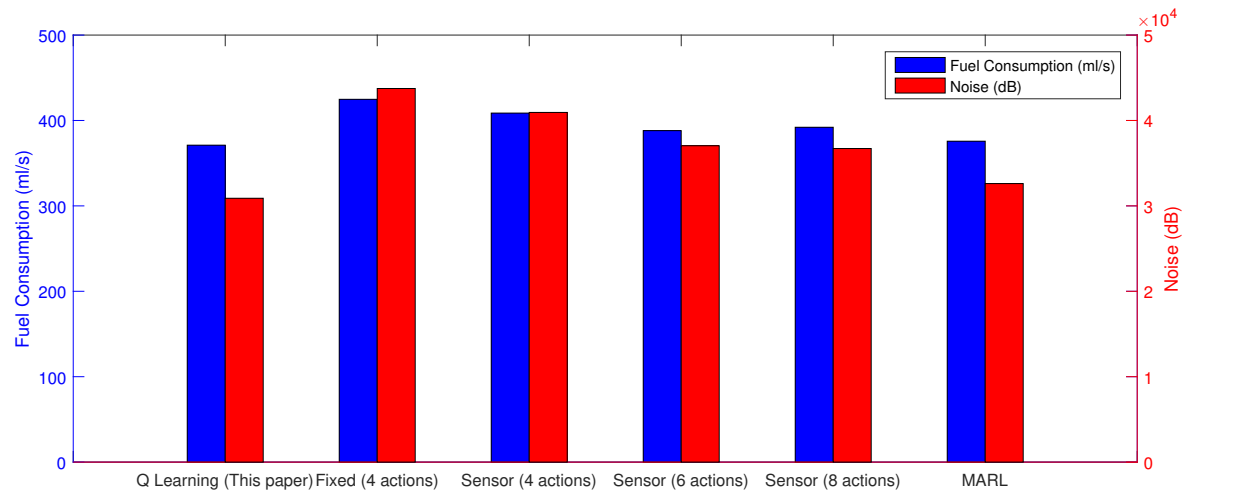

Fig. 7: Fuel consumption and noise pollution of vehicles in the case of high pedestrian rate

Figures 4-7 show the simulation results in the case of high pedestrian rate. Figure 4 shows that total cumulated queue length for vehicles and pedestrians. Note that the total queue length in Fig. 4 represents the total number of vehicles or pedestrians that are waited at all the intersections due to red lights. If a vehicle or a pedestrian is not waited at an intersection (e.g. green lights), the vehicle or pedestrian is not taken into account when the total queue length is calculated at this intersection. It can be seen that, compared with the fixed-time and sensor-based dynamic control solutions, the Q learning based solutions, including both the proposed solution in this paper and the MARL solution can greatly reduce the vehicle and pedestrian queue lengths. It is because of the nature of reinforcement learning which is capable of optimizing the control actions, and thus, more vehicles or pedestrians do not wait at intersections, resulting in smaller total queue length. If we compare the performance between the Q learning proposed in this paper and MARL, we can observe that our solution is better than MARL. The major reason is that our solution considers both the vehicle and pedestrian information, while MARL only considers vehicles. Moreover, our solution exchanges information among multiple Q learning agents, which is useful to achieve the optimization for the entire system. Through the GUI of SUMO, we observed that our solution can achieve a smoother traffic flow during the simulation, compared with other solutions. Figure 5 shows the waiting and traveling time comparison results. Similarly, our solution outperforms others. Among the results, the vehicle and pedestrian waiting time is the most straightforward performance metric because an efficient

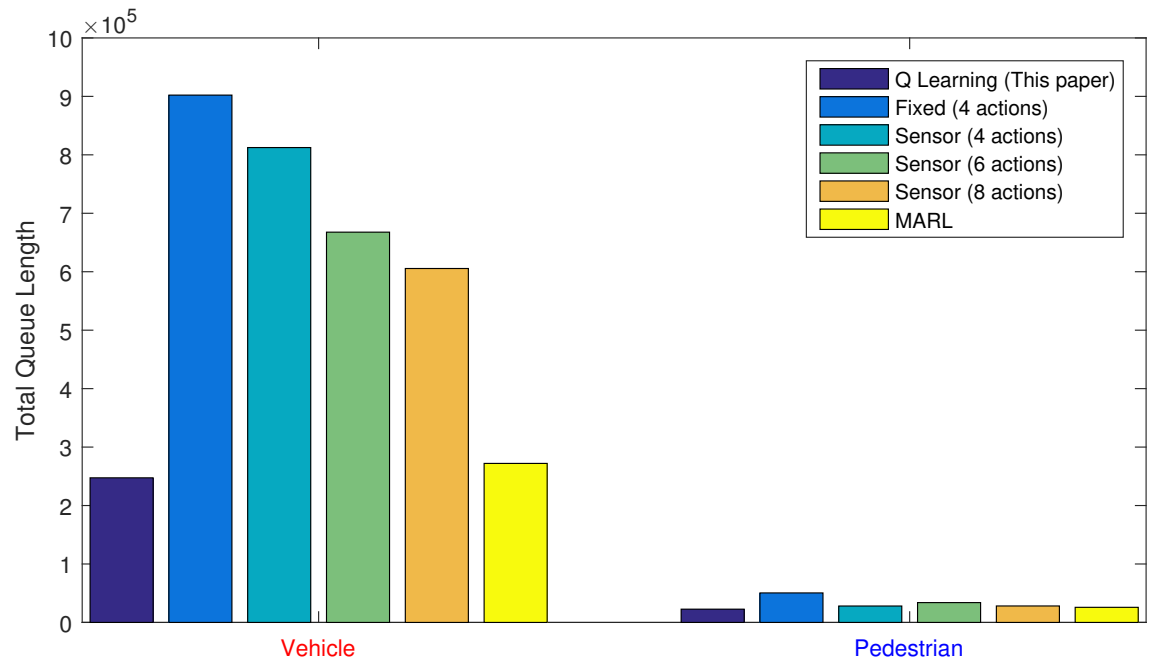

Fig. 8: Total cumulated queue lengths for vehicles and pedestrians in the case of medium pedestrian rate

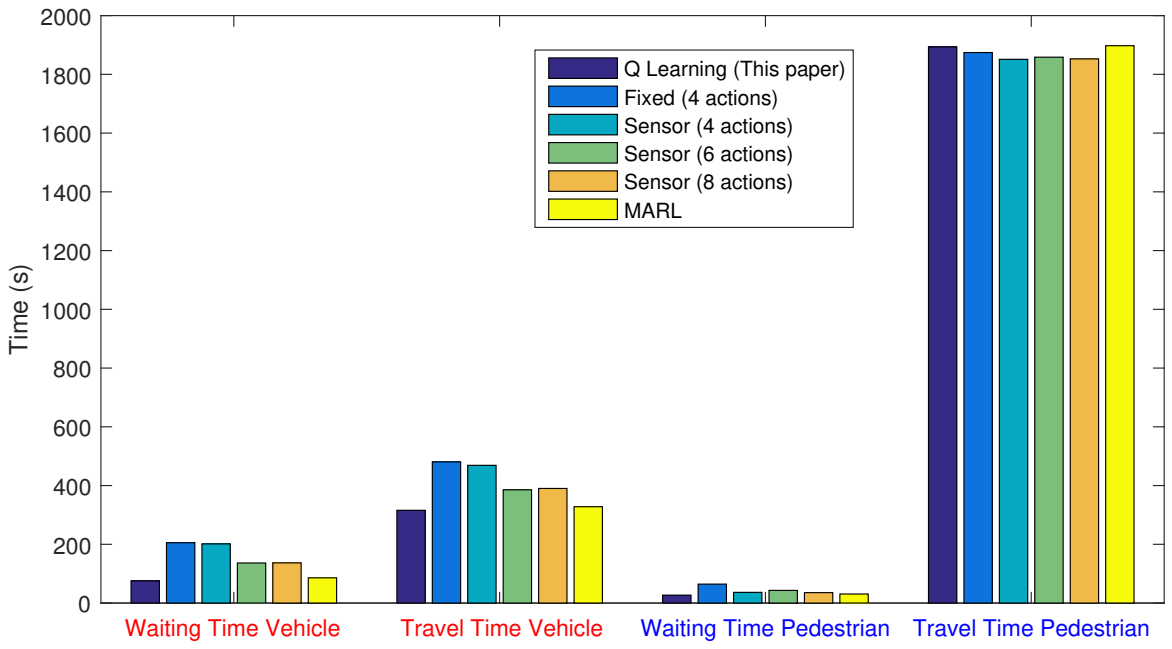

Fig. 9: Total waiting and traveling time for vehicles and pedestrians in the case of medium pedestrian rate

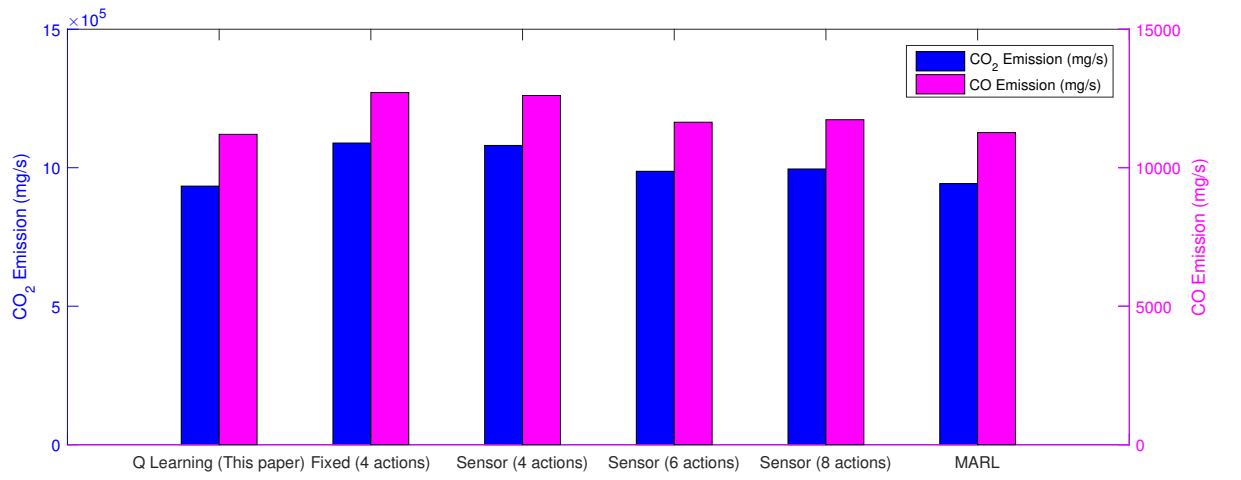

Fig. 10: $CO_2$ and $CO$ emission of vehicles in the case of medium pedestrian rate

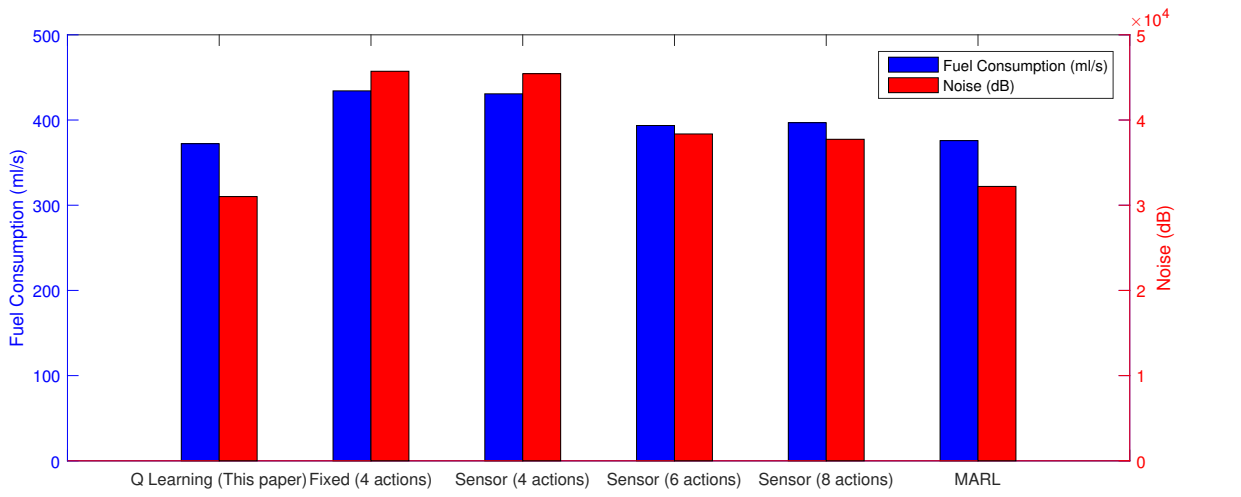

Fig. 11: Fuel consumption and noise pollution of vehicles in the case of medium pedestrian rate

traffic light control algorithm should always reduce the waiting time of vehicles and pedestrians at intersections. Due to the shorter waiting time and reduction of unnecessary stopping and starting of traffic, our solution can, in turn, reduce $CO_2$ and $CO$ emissions, fuel consumption, and noise pollution of vehicles, as depicted in Figs. 6 and 7 respectively. From all the results depicted in Figs. 4-7, we can see that the proposed distributed multi-agent Q learning is better than exiting solutions and MARL in term of many key performance metrics, which validated the overall feasibility and efficiency of the proposed solution in terms of traffic light control.

Figures 8-11 show the simulation results in the case of medium pedestrian rate, and Figures 12-15 show the simulation results in the case of low pedestrian rate. In detail, Figures 8, 9, 10 and 11 depict the simulation results regarding vehicle and pedestrian queue lengths, vehicle and pedestrian waiting and travelling time, $CO_2$ and $CO$ emission, and fuel consumption and noise pollution in the case of medium pedestrian rate respectively. Figures 12, 13, 14 and 15 depict the simulation results in terms of vehicle and pedestrian queue lengths, vehicle and pedestrian waiting and travelling time, $CO_2$ and $CO$ emission, and fuel consumption and noise pollution in the case of low pedestrian rate respectively. From the results, we can see that, similar to the case with high pedestrian rate, our proposed solution also outperforms other solutions when the pedestrian rate is medium or low. In other words, the proposed solution performs best in all the simulated scenarios, which validated that the proposed solution can be deployed in different type of cities with different traffic patterns. If we further compare the results for the high, medium and low pedestrian rates, we can observe that our solution has larger performance improvement compared with MARL when there are more pedestrians. Take the vehicle waiting time as examples, in the case of high pedestrian rate, our solution can achieve 16.7% improvements for vehicle waiting time reduction compared with MARL. In the case of medium and low pedestrian rates, the number is reduced to 12.2% and 7.0% respectively. This observation indicates that our solution is more efficient than MARL when there are more pedestrians. It is reasonable because our solution jointly considers the vehicle and pedestrian traffic for optimization while MARL only considers the vehicle traffic.

## VI. Conclusions and Future Works

This paper presents an intelligent traffic light control system which takes pedestrians into account in order to achieve optimization for both motorized and non-motorized traffic. The system is empowered by a distributed multi-agent Q learning, which is able to collaboratively calculate the optimal control actions, based on the traffic information not only from the local intersection, but also from neighboring intersections. Moreover, many real-world constraints / rules for traffic light control are integrated in the Q learning algorithm, which can facilicate the proposed solution to be deployed in real operational scenarios. Numerical simulations are carried out based on a real-world map with real-world traffic data. The simulation results show that our proposed solution outperforms existing solutions in terms of vehicle and pedestrian queue length, waiting time at intersections, and many other key performance metrics such as emissions and fuel consumptions.

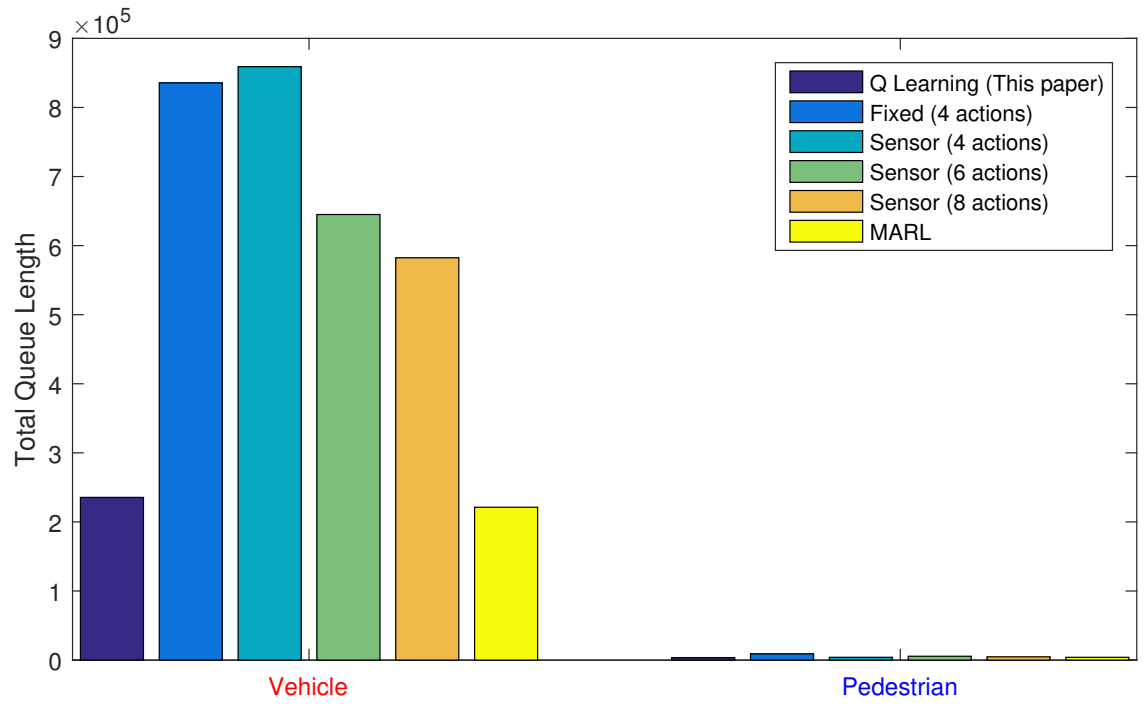

Fig. 12: Total cumulated queue lengths for vehicles and pedestrians in the case of low pedestrian rate

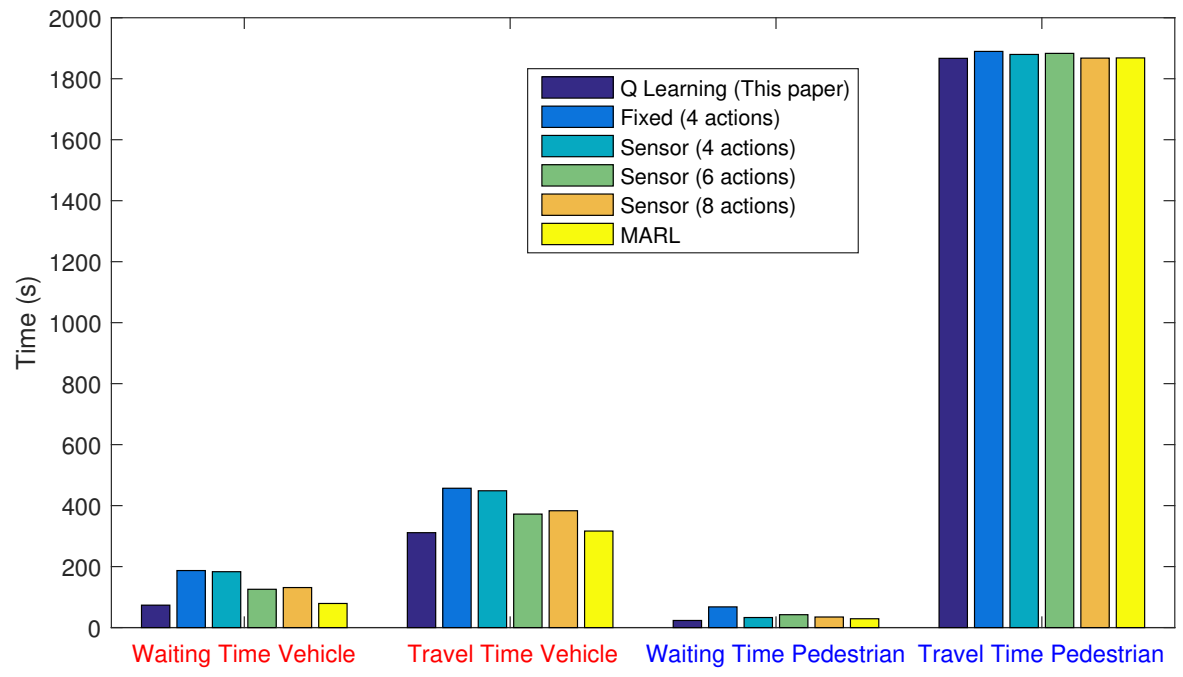

Fig. 13: Total waiting and traveling time for vehicles and pedestrians in the case of low pedestrian rate

Our future works are twofold. Firstly, we will further improve the algorithm performance, for example, the Q learning convergence time, in order to handle the scenario where sharp change of traffic pattern occurs. Secondly, we will investigate different deployment models for the proposed system, in addition to the fully distributed model used in ths paper, to evaluate whether the performance of the entire system can be further improved. As the AI and machine learning technology has proven to be useful in many use cases, we hope the work presented in this paper can shed light on the future real deployment of AI based traffic light control system.

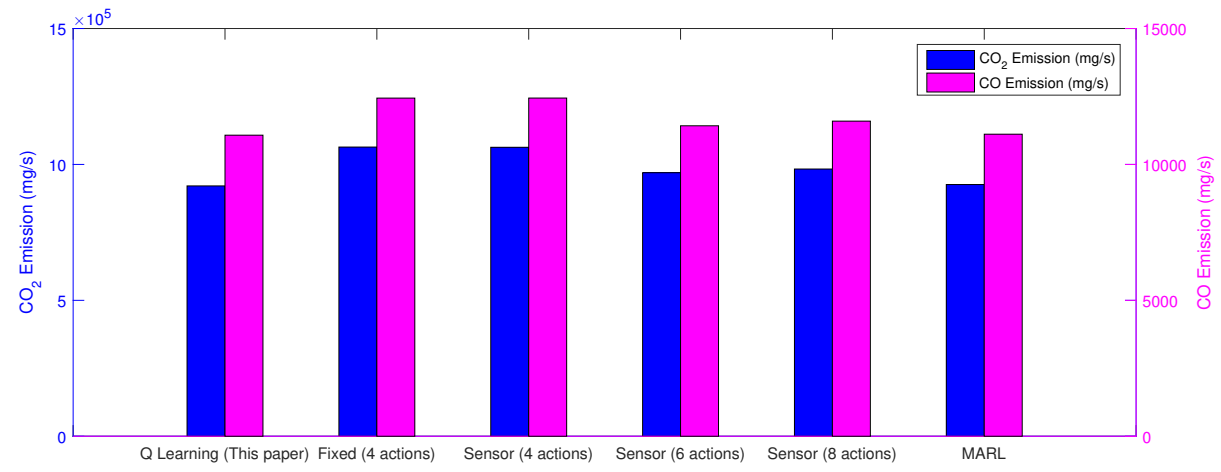

Fig. 14: $CO_2$ and $CO$ emission of vehicles in the case of low pedestrian rate

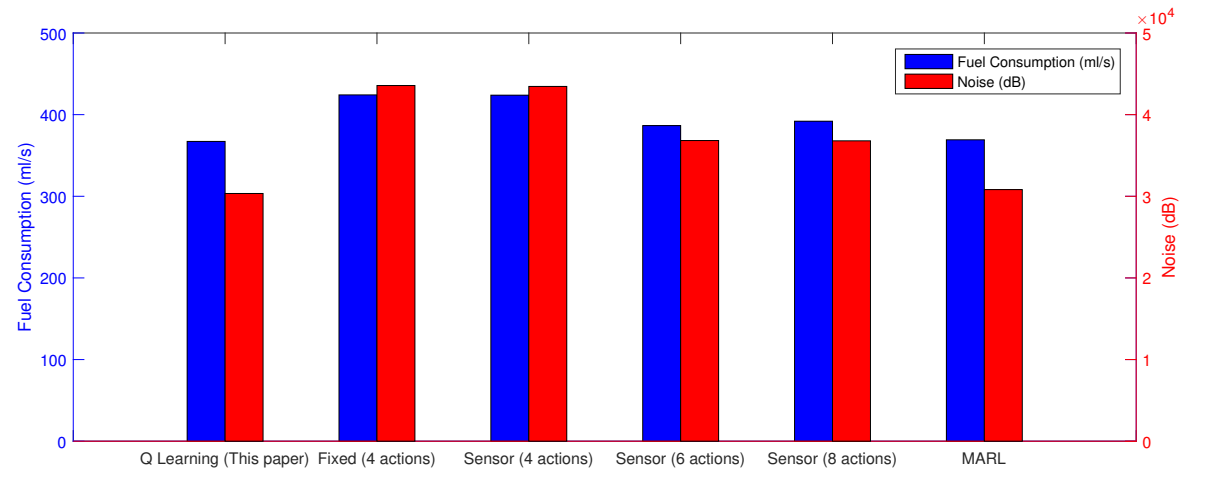

Fig. 15: Fuel consumption and noise pollution of vehicles in the case of low pedestrian rate